\documentclass[aps,prl,twocolumn,superscriptaddress,showpacs]{revtex4}
\usepackage{float}
\usepackage{graphicx}
\begin{document}

\title{Nonlinear elasticity of composite networks of stiff biopolymers with
flexible linkers}

\author{C.P. Broedersz}
\affiliation{Department of Physics and Astronomy,
Vrije Universiteit, 1081 HV Amsterdam, The
Netherlands}
\author{C. Storm}
\affiliation{Department of Physics and Astronomy,
Vrije Universiteit, 1081 HV Amsterdam, The
Netherlands} \affiliation{Department of Applied
Physics and Institute for Complex Molecular
Systems, Eindhoven University of Technology, P.
O. Box 513, NL-5600 MB Eindhoven, The
Netherlands}
\author{F.C. MacKintosh}
\email{fcm@nat.vu.nl} \affiliation{Department of
Physics and Astronomy, Vrije Universiteit, 1081
HV Amsterdam, The Netherlands}

\date{\today}

\begin{abstract}
%Motivated by recent experiments showing striking nonlinear
%elasticity of in vitro networks of the biopolymer actin cross-linked
%with filamin, we present an effective medium theory of flexibly
%cross-linked stiff polymer networks. We model such composite
%networks as a collection of randomly oriented elastic rods, each of
%which is mechanically connected by flexible connectors to a
%surrounding elastic continuum, which self-consistently represents
%the behavior of the rest of the network. This model yields a
%cross-over from a linear elastic regime to a highly nonlinear
%elastic regime that stiffens in a way quantitatively consistent with
%experiment.
Motivated by recent experiments showing nonlinear
elasticity of in vitro networks of the biopolymer
actin cross-linked with filamin, we present an
effective medium theory of flexibly cross-linked
stiff polymer networks. We model such networks by
randomly oriented elastic rods connected by
flexible connectors to a surrounding elastic
continuum, which self-consistently represents the
behavior of the rest of the network. This model
yields a cross-over from a linear elastic regime
to a highly nonlinear elastic regime that
stiffens in a way quantitatively consistent with
experiment.
\end{abstract}

%\keywords{}

\maketitle The mechanical response of living cells depends largely
on their \emph{cytoskeleton}, a network of stiff protein polymers
such as F-actin, along with various associated proteins for
cross-linking and force generation. In addition to their importance
for cell mechanics, cytoskeletal networks have also demonstrated
novel elastic properties, especially in numerous \emph{in vitro}
studies~\cite{JanmeyJCB91, Gardel04, Storm05, BauschNatPhys,
Chaudhuri07, Janmey07}. The cellular cytoskeleton, however, is an
inherently composite structure, consisting of elements with highly
varied mechanical properties, and there have been few theoretical or
experimental studies of this aspect~\cite{Gardel06,
Wagner06,Kasza2007, DiDonna06, DischerLubensky}. Recent experiments
on F-actin with the physiological cross-linker filamin have
%Comm_deletion:demonstrated several striking features of such networks: while their
demonstrated several striking features while
their linear modulus is significantly lower than
for rigidly cross-linked actin systems, they can
nonetheless withstand remarkably large stresses
and can stiffen by a factor of 1000 with applied
shear~\cite{Gardel06,Kasza2007,Karen}. This
behavior appears to result from the highly
flexible nature of filamin, although the basic
physics of such a network, in which the
elasticity is dominated by cross-linkers, is not
understood. Apart from their physiological
importance, such networks suggest new principles
that may be extended to new synthetic materials
with designed cross-links~\cite{Wagner06}.

Here, we develop a theoretical model for composite networks of rigid
filaments connected by flexible cross-linkers, in which the
macroscopic network elasticity is governed by the cross-links. We
examine this model in a limit in which the basic elastic element is
a single rigid rod, directly linked by numerous compliant
cross-linkers to a surrounding linear elastic medium. We show that
such a network stiffens in a manner determined by the mechanics of
individual cross-links, which we model both as linear springs with
finite extension, and also as wormlike chains. We analyze our model
in both a fully 3D network, as well as a simplified 1D
representation, which already captures the essential physics of the
nonlinear behavior. The finite extension $\ell_0$ of the cross-links
along with the length of the filaments/rods $L$ implies that there
exists a characteristic strain $\gamma_c\sim\ell_0/L$ for the onset
of the nonlinear response of the network. Indeed prior \emph{in
vitro} experiments, in which the length of the cross-linkers was
varied~\cite{Wagner06}, have reported this linear dependence on
$\ell_0$. We extend this model in a fully self-consistent manner,
replacing the embedding medium by an effective medium whose elastic
properties are determined by those of the constituent rods and
linkers. This self-consistent model can quantitatively account for
the nonlinear response found in prior experiments on actin-filamin
networks \cite{Gardel06,Kasza2007}.

\begin{figure}
\centering
\includegraphics[width=180 pt]{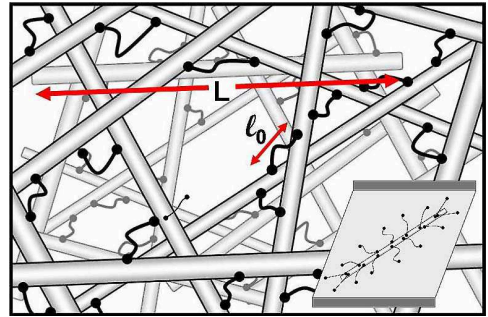}\vspace{-.in}
\caption{(Color online) Schematic figure of an
isotropic stiff polymer network with highly
compliant cross-linkers. The inset illustrates
the proposed non-uniform deformation of the
cross-linkers on a single filament in a sheared
background medium.} \label{FlexConNetwork}
\end{figure}
\begin{figure}
\centering
\includegraphics[width=180 pt]{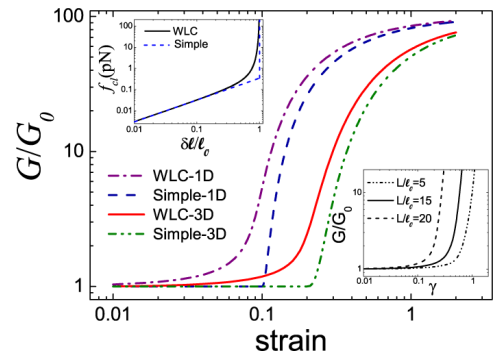}\vspace{-.0in}
\caption{(Color online) The shear modulus $G$ normalized by the
linear modulus $G_0$ as a function of strain $\epsilon$ (1D) or
$\gamma$ (3D) for simple cross-linkers and WLC cross-linkers in the
1D and the 3D version of the linear medium hairy rod model. In this
plot we have chosen $K_{EM}=100 k_{cl}$ as an example. The upper
left inset shows the force extension curve of a simple cross-linker
(blue dashed curve) and of a WLC cross-linker (black solid curve).
The lower right inset shows the normalized shear modulus as a
function of strain $\gamma$ for various ratios of $L/\ell_0$
calculated with the self-consistent model.} \label{PhenMod}
\end{figure}

 In a flexibly cross-linked stiff polymer network, randomly oriented stiff filaments or rods are
interconnected by relatively short but highly flexible cross-linkers
(Fig.~\ref{FlexConNetwork}). The compliance of this network is
dominated by the flexible cross-linkers, while the much
stiffer filaments act mainly as a scaffold for the cross-linkers,
ensuring rigidity of the network as a whole. Recent experiments have
demonstrated that flexible biological cross-linkers such as filamin
can be described as a semiflexible polymer using the wormlike chain
(WLC) model~\cite{Schwaiger,Furuike01}. The cross-linkers are
characterized by their contour length $\ell_0$ and persistence
length $\ell_p$~\cite{Bustamante}. A realistic force-extension curve
of a typical biological cross-linker is shown as a black solid line
in the inset of Fig.~\ref{PhenMod}. It is instructive to simplify
this curve by assuming linear response with a spring constant
$k_{cl}$ and a finite extensibility $\ell_0$. This simplification
retains the essential features,
and is shown in Fig.~\ref{PhenMod} as a blue dashed line. We refer
to this as \emph{simple} cross-link behavior.

To determine the elasticity of the network we use
an effective medium approach, and divide the
network into two mechanically connected
sub-systems. The first consists of a stiff
%COMM_replacement:filament of length $L$ decorated by many flexible
filament of length $L$ decorated by $n$ flexible
cross-linkers, which we refer to as a hairy rod
(HR). The other is the network connected to it,
which we treat as an elastic continuum. %The
%COMM_deletion:composite network can be modeled by a collection
%of independent, randomly oriented rigid rods that
%are each connected to the medium by $n$ flexible
%cross-linkers.
Although the medium is assumed to deform
affinely, we allow the local strain of the
cross-linkers to depend on their position on the
HR. By averaging over all orientations we may
express the macroscopic stress in terms of the
tension in a single HR connected to a medium,
which is subject to a 1D strain $\epsilon$ along
its backbone. The tension $\tau$ in the center of
this HR is the sum of the forces exerted by all
cross-linkers on one half of the rod. To
calculate these forces we treat the cross-linker
as a spring connected in series with the medium,
which we describe with a spring constant
$K_{EM}$. We are primarily interested in densely
cross-linked networks for which $K_{EM} \gg
k_{cl}$. The extension of the cross-linker-medium
system is given by $\epsilon x$ at a distance $x$
from the center of the rod. If the cross-linkers
are homogenously distributed over the rod with a
high density $n/L$ we can write the sum over
forces as an integral,
\begin{eqnarray}
\label{TensionSimple}
    \tau(\epsilon)&=& \frac{n}{L} \int_0^{\frac{\ell_0}{\epsilon}} \, dx \, \frac{k_{cl} K_{EM}}{k_{cl}+K_{EM}}\epsilon x\\
    &+&\frac{n}{L} \int_{\frac{\ell_0}{\epsilon}}^\frac{L}{2} \, dx \,
    \left[\frac{k_{cl} K_{EM}}{k_{cl}+K_{EM}}\ell_0+K_{EM}(\epsilon x-\ell_0)\right]\nonumber
\end{eqnarray}
For strains $\epsilon \leq \frac{\ell_0}{L/2}$ only the first
integral is present and the integration extends to $L/2$. In this
case, the tension depends linearly on the strain. Using
Eq.~(\ref{TensionSimple}) we compute the 1D modulus
$G_{1D}=\tau/\epsilon$, which is shown as a dashed blue line in
Fig.~\ref{PhenMod}. For small strains the system is linearly elastic
with $G_{1D}=\frac{1}{8} n \frac{k_{cl} K_{EM}}{k_{cl}+K_{EM}} L $.
Above a threshold strain $\frac{\ell_0}{L/2}$ a cross-over occurs to
a second linear regime in which $G_{1D}$ asymptotically approaches
$\frac{1}{8} n K_{EM} L$.

%Comm_deletion:The nonlinear response of a cross-linker has an
%entropic origin and
The nonlinear response of a cross-linker is more
realistically modeled with the WLC
model~\cite{Furuike01} (Fig.~\ref{PhenMod}). We
calculate the tension in a rod with WLC
cross-linkers analogously to
Eq.~(\ref{TensionSimple}). The 1D modulus
$G_{1D}$ is shown as a purple dash-dotted line in
Fig.~\ref{PhenMod}. Though quite similar to the
simple cross-linker model, the more realistic
force-extension curve has introduced a
considerable smoothing of the cross-over
resulting in a gradual onset of nonlinear
behavior of the HR with WLC cross-linkers.
Nevertheless, the characteristic strain
$\epsilon_c$ for the nonlinear behavior is
proportional to $\ell_0/L$ independent of the
exact nonlinear response of the linkers.

Using the 1D model presented above we can compute the macroscopic
stress of a network. A 3D isotropic network with a polymer length
density $\rho$ is modeled by an
effective medium consisting of randomly oriented
HR's. We can compute the macroscopic stress $\sigma$ and shear
modulus $G=\sigma/\gamma$ by averaging over all orientations~\cite{Gardel04,Janmey07}. The shear
modulus is shown in Fig.~\ref{PhenMod} for the simple cross-linkers
and for the WLC cross-linkers. The 3D curves are largely similar to the 1D results, save for a factor two shift which may be understood by noting that the rods
at a $45^\circ$ angle to the stress plane, which bear
most of the stress, experience an extensional strain $\epsilon$ of
$\gamma/2$.

%{\em Self-Consistent Medium Approach}\\
At large strains, when many of the cross-linkers are extended well
into their nonlinear regimes, it is no longer realistic to assume
a linear background medium. To address this, we shall now require the elasticity of the background medium to {\em self-consistently}
represent the nonlinear elasticity of its constituent HR's.

Under strain the cross-links deform the
surrounding elastic medium. The resulting
longitudinal displacement $\delta\ell$ of the
medium leads to a restoring force per unit length
along the rod given approximately by the shear
stiffness $\frac{d\sigma}{d\gamma} \times
\delta\ell$
\endnote{The longitudinal elastic restoring force per unit length of
a rod with aspect ratio $L/a$ is approximately $2\pi
G/\log(L/a)\delta\ell$; here we ignore the log term.} The shear
modulus of the medium depends on the density $\rho$ of rods and the
longitudinal stiffness $\frac{d\tau}{d\epsilon}$, where $\epsilon$
is the 1D extensional strain of the medium along the rod:
\begin{equation}\label{SelfConModulus}
   \frac{d\sigma}{d\gamma}=\mathcal{A}\rho\frac{d\tau}{d\epsilon}.
\end{equation}
Here, $\mathcal{A}$ is a dimensionless geometric factor that depends
on the architecture of the network. For an isotropic network in 3D,
this is 1/15. (We note that this is a small-strain approximation and
that $\mathcal{A}$ will be different for anisotropic networks.)
Thus, the effective stiffness $K_{EM}$ per cross-link is given by
\begin{equation}\label{SelfConReq}
    K_{EM}=\mathcal{A} \rho \frac{L}{n} \frac{d\tau}{d
    \epsilon}.
\end{equation}
When subject to a shear strain $\gamma$, the resulting stress
$\sigma$ within a network of rods can be expressed in terms of the
tension $\tau$ in each rod, which depends on its orientation
relative to the shear plane. It is given by
\begin{equation}
\label{TensionSC}
   \tau(\epsilon) = \frac{n}{L} \int_0^{\frac{L}{2}} \, dx'
   \,x'
   \int_0^\epsilon \, d\epsilon' \frac{k_{cl}(x' \epsilon') \mathcal{A} \rho \frac{L}{n} \frac{d \tau}{d
   \epsilon}(\frac{x'
    \epsilon'}{L/2})}{k_{cl}(x' \epsilon')+\mathcal{A} \rho \frac{L}{n} \frac{d \tau}{d
    \epsilon}(\frac{x'
    \epsilon'}{L/2})
   }
\end{equation}
where $k_{cl}(\delta \ell)$ is the derivative of
the force-extension curve of the cross-linker.
Equivalently, we may write for
$\tau(\epsilon)$
\begin{eqnarray}
\label{SelfConWLCDE}
   2 \frac{d\tau}{d\epsilon}&+&\epsilon
\frac{d^2\tau}{d\epsilon^2} = \\
&&\left\{
\begin{array}{ll}
    \frac{n L}{4} \frac{k_{cl}(\epsilon L/2) \mathcal{A} \frac{\rho L}{n}\frac{d\tau}{d\epsilon}}{k_{cl}(\epsilon L/2)+\mathcal{A} \frac{\rho L}{n}
\frac{d\tau}{d\epsilon}} & \textrm{if $\epsilon < \frac{l_0}{L/2}$}\\
    \\
    \frac{1}{4} \mathcal{A} \rho L^2 \frac{d\tau}{d\epsilon}
     & \textrm{if $\epsilon \geq\frac{l_0}{L/2}$}
    \end{array} \right.\nonumber
\end{eqnarray}
%Comm_deletion:It is instructive to first investigate the
%properties of this model
We first investigate the properties of this model
using the simple force-extension curve (see inset
Fig.~\ref{PhenMod}). For a densely cross-linked
network we find a linear regime below
$\gamma_c=\frac{\ell_o}{L/2}$. For larger strains
the system enters a highly nonlinear regime for
which
\begin{equation}\label{exponent}
     \frac{d\tau}{d\epsilon} \sim \tau^{1-1/ (\frac{1}{4}\mathcal{A} \rho
     L^2-1)}.
\end{equation}
This is in marked contrast with the linear medium model in which
there is only a cross-over between two distinct linear regimes.

\begin{figure}
\centering
\includegraphics[width=180 pt]{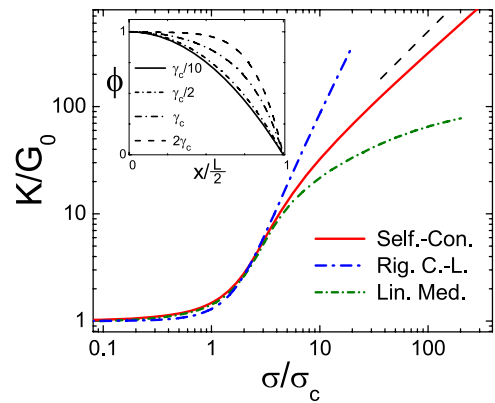}\vspace{-.0in}
\caption{(Color online) Differential modulus $K={d\tau}/{d\gamma}$
normalized by the linear modulus $G_0$ as a function of stress
$\sigma$ normalized by $\sigma_c$ for the self-consistent
(Self.-Con.) model. We also plot $K/G_0$ for the linear medium (Lin.
Med.) model and a model for rigidly cross-linked semiflexible
polymer networks (Rig. C.-L.). The black line indicates a slope of
1. The inset shows the reduced tension profile $\phi$ along the rod,
normalized by the mid-point tension $\tau$ in Eqs.~(\ref{TensionSC})
and (\ref{SelfConWLCDE}).} \label{selfcon}
\end{figure}

A real network with compliant cross-linkers is more realistically
modeled by solving Eq.~(\ref{SelfConWLCDE}) (numerically) using the
WLC force-extension curve for the cross-linkers.
The shear modulus in this case, computed exactly as before, is graphed in the lower right inset
of Fig.~\ref{PhenMod}. At low strains $G\sim n k_{cl} \rho L$ and
there is a gradual onset of nonlinear response originating from the
nonlinear entropic elasticity of the cross-linkers(see upper left
inset Fig.~\ref{PhenMod}). At a strain $\sim \ell_0/L$ the
cross-linkers at the edges of the rods become effectively rigid,
which marks the onset of the nonlinear network behavior.

In view of the nonlinearity of this system, it is more appropriate to use
a differential modulus $K=\frac{d\sigma}{d\gamma}$ rather than $G$.
The differential modulus is plotted in
Fig.~\ref{selfcon}. Up to a critical stress $\sigma_c$, the elasticity is dominated by WLC
cross-linkers placed on a rigid rod connected to a much stiffer
medium. At larger stresses, the cross-linkers at the edges of the HR
reach full extension and, consequently, couple strongly to the
surrounding network. In this limit, the slope in a $\log(K)$ vs
$\log(\sigma)$ plot approaches $\approx 1-1/(\frac{1}{4}\mathcal{A}
\rho L^2-1)$, as it does for simple cross-linkers.
This exponent is a consequence of the composite nature of the
network and its nonlinear constituents, although it is independent
of the exact form of the nonlinear response of the cross-linkers.
For a dense flexibly cross-linked network $\rho L^2\gg1$ and,
therefore, we expect a slope of 1. This is consistent with recent
experimental data on actin networks cross-linked by the highly
compliant cross-linker filamin in which a slope of 1 was
found~\cite{Gardel06} in contrast to a slope of 3/2 found for
rigidly cross-linked networks~\cite{Gardel04,Kasza2007}.
Interestingly, \emph{in vivo} experiments show that cells also
exhibit powerlaw stiffening with an exponent of
1~\cite{Wang,Fernandez}.

We compare our results to the linear medium model and a model based on the nonlinear response of the semiflexible
actin segments between cross-links that has been used successfully
to describe rigidly cross-linked actin
networks~\cite{Gardel04, MacKPRL95} in Fig.\ \ref{selfcon}. Although
the three curves coincide for small stresses, at intermediate
stresses $\sigma \gtrsim \sigma_c$ the linear medium model curve
rolls over to a linear regime. Clearly, our self-consistent model and the model for rigidly linked networks begin to differ in the nonlinear regime.

So far, we have considered only the mid-point tension. In networks
of elastic filaments of finite length, however, the tension along a
single filament is not uniform, but decreases towards its
ends~\cite{Morse98,Head03,Heussinger07}. The inset of Fig.\
\ref{selfcon} shows the ratio $\phi$ of the tension at point $x$
along the rod to the maximum tension. This maximum occurs at the
mid-point $x=0$, and is given by $\tau$ in Eqs.~(\ref{TensionSC})
and (\ref{SelfConWLCDE}). The tension at a point $x$ can be obtained
by replacing the lower limit of of the $x'$ integral in
Eq.~(\ref{TensionSC}) by $x$. The tension profile is parabolic below
$\gamma_c$ and quickly converges to a more flattened out profile in
the nonlinear regime. We can use the tension profile to relate the
maximum tension in a single HR to the macroscopic stress
$\sigma$~\cite{Broederszlongpaper}. For typical experimental
conditions in an actin-filamin gel\cite{Karen} we estimate a maximum
force on a single cross-link to be at most $5\ \text{pN}$ for
isotropic rods and of order $1\ \text{pN}$ or less for oriented
rods.

A feature shared by the linear medium model and the self-consistent
model is the characteristic strain $\gamma_c\simeq4\ell_0/L$ for the
onset of nonlinear response. The
proportionality with $\ell_0$ is consistent with the results of
Wagner et al.\, where cross-linker length was varied, although they
observed larger values of $\gamma_c$ than expected either from our
model and based on Refs.\cite{Gardel06,Kasza2007}. Recent
experiments on actin-filamin networks also show a dependence of the
critical strain that is approximately inverse in actin filament
length $L$~\cite{Liu} in agreement with our results.
This sensitivity of network response to filament length, both in experiments and in our model, appears to be one of the hallmarks of actin-filamin networks.
On the one hand, this may explain the apparent difference between the critical strains reported in Refs.\ \cite{Wagner06,Gardel06,Kasza2007}.
On the other hand, it also suggests that it may be more important to directly measure the filament length distribution in such experiments than in other similar \emph{in vitro} studies.
In Wagner et al., for instance, the filament length was not measured, but was inferred from prior reports of the length dependence on the capping protein gelsolin \cite{PaulGelsolin}.

In previous work, DiDonna and Levine have assumed
a sawtooth force-extension curve for the
cross-linkers to mimic domain unfolding. They
report a fragile state with shear softening when
an appreciable number of cross-linkers are at the
threshold of domain unfolding~\cite{DiDonna06}.
Our model is based on the stiffening of the
cross-linkers, which occurs at forces far below
those required for domain
unfolding~\cite{Schwaiger, Furuike01}. This leads
to strain stiffening at a point where only a
fraction of cross-linkers are at their threshold
for nonlinear response. Thus in both our model
and that of Ref.\ \cite{DiDonna06} the network
responds strongly to small strain changes, though
in an opposite manner: stiffening in the present
case vs. softening in Ref.\ \cite{DiDonna06}. In
related work, Dalheimer et al.\ show that
isotropic networks linked by large compliant
cross-linkers exhibit a shear induced ordering
transition to a nematic
phase~\cite{DischerLubensky}. Our model accounts
for the architecture of the network through an
averaging procedure in a scalar quantity
$\mathcal{A}$. We are presently investigating the
effect of an ordering transition on the nonlinear
response of the network.

We have introduced a model for flexibly cross-linked stiff polymer
networks based on cross-linker elasticity. Our model yields an
exponent of $~1$ in the asymptotic powerlaw behavior of a $K$ vs
$\sigma$ curve in agreement with experiments on \emph{in vitro}
filamin-actin networks\cite{Gardel06,Kasza2007}. The exact form of
the nonlinear response predicted by our model can be tested by
further experiments\cite{Karen}.

\begin{acknowledgments}
We thank G.\ Koenderink and K.\ Kasza for useful discussions. This
work was funded in part by FOM/NWO.
\end{acknowledgments}

\end{document}